\documentclass[prd,showpacs,twocolumn,superscriptaddress,floatfix,10pt]{revtex4-2}

\usepackage{setspace} 
\usepackage[utf8]{inputenc}
\usepackage{float}
\usepackage{amsmath,amssymb,amsfonts,bm}
\usepackage{graphicx}
\usepackage{subfigure}
\usepackage{multirow}
\usepackage[usenames,dvipsnames]{color}
\allowdisplaybreaks
\usepackage[
colorlinks=true,
linkcolor=blue,
breaklinks=true,
urlcolor=blue,
citecolor=blue]{hyperref}

\usepackage[sort&compress]{natbib}
\usepackage{comment}
\graphicspath{{./figs/}}
\usepackage{epstopdf}
\usepackage{appendix}
\usepackage{pstricks}
\usepackage{epsfig}
\usepackage{cancel}
\usepackage{braket}
\usepackage[normalem]{ulem} 
\usepackage{tikz-feynman}
\usepackage{orcidlink}

\newcommand{\be}{\begin{equation}}
\newcommand{\ee}{\end{equation}}
\newcommand{\ben}{\begin{eqnarray}}
\newcommand{\een}{\end{eqnarray}}

\usepackage{xcolor}
\newcommand{\vek}{{\bm k}}

\newcommand{\UFBA}{Instituto de Física, Universidade Federal da Bahia, Campus Ondina, Salvador, Bahia 40170-115, Brazil}

\newcommand{\UFPI}{Departamento de Física, Universidade Federal do Piauí, Teresina, Piauí 64049-550, Brazil}

\begin{document}

% ---------------------------------------------------------------------
% TITLE AND AUTHORS (remove math from PDF sting with texorpdfstring)
% ---------------------------------------------------------------------

\title{A pion-driven near-threshold enhancement in the $\pi\,T_{cc}$ system}

\author{Pedro Brandão\orcidlink{0000-0002-1470-184X}}
\email[]{pedro.brandao@ufba.br}
\affiliation{\UFBA}

\author{Jorgivan Morais Dias\orcidlink{0000-0002-0354-4711}}
\email[]{jorgivan.dias@ufpi.edu.br}
\affiliation{\UFPI}

\author{Luciano M. Abreu\orcidlink{0000-0001-7408-1913}}
\email[]{luciano.abreu@ufba.br}
\affiliation{\UFBA}

\begin{abstract}

Hadronic molecules are commonly viewed as products of near-threshold two-body dynamics. 
We investigate whether an experimentally established molecule can also serve as a building 
block for a more complex exotic hadron by studying pion scattering off $T^+_{cc}(3875)$. 
Treating $T_{cc}$ as a correlated isoscalar $D D^*$ molecular cluster, we describe the 
resulting $\pi D D^*$ dynamics in the $I(J^P)=1(1^-)$ channel within the Fixed Center Approximation to the Faddeev equations. 
The full three-body amplitude develops a pronounced enhancement slightly above the $\pi T_{cc}$ 
threshold, centered at $M_{\rm peak}\simeq 4055$~MeV, with an apparent width of approximately $60$~MeV. The associated behavior of the real and imaginary parts of the amplitude is compatible with a resonant-like interpretation. These results show that the 
lightest hadron can play an active dynamical role in promoting an observed hadronic molecule 
into a building block of a heavier near-threshold structure generated by the three-body dynamics. 
We propose searching for this enhancement in the $T_{cc}\pi$ invariant-mass distribution, 
particularly through the $D^0D^0\pi^+\pi^-$ final state.

\end{abstract}

\maketitle
\newpage

% ---------------------------------------------------------------------
% ---------------------------------------------------------------------
% ---------------------------------------------------------------------
%{\it Introduction.}---
% ---------------------------------------------------------------------
% ---------------------------------------------------------------------
% ---------------------------------------------------------------------

{\it Introduction.} Near-threshold exotic hadrons provide a particularly 
sensitive probe of the long-distance regime of quantum chromodynamics. Many 
prominent candidates are naturally interpreted as hadronic molecules, namely, 
states generated predominantly by residual strong interactions between color-singlet 
hadrons \cite{Guo:2017jvc,Liu:2024uxn,Wang:2025dur}. In this picture, their 
proximity to a two-hadron threshold is not incidental but reflects the dynamical 
origin of the state. A broader few-body question then arises: once such a molecular 
state is formed, can it itself become a building block for a more complex hadronic 
structure?

The observation of the doubly charmed $T^+_{cc}$ by the LHCb Collaboration 
makes this question especially timely~\cite{LHCb:2021auc,LHCb:2021vvq}. The state 
was identified as a narrow structure in the $D^0D^0\pi^+$ spectrum, with a mass 
extremely close to $D^{* +} D^0$ threshold. Its near-threshold location has motivated 
extensive studies of $T_{cc}$ as a shallow isoscalar $D D^*$ molecule 
~\cite{Ling:2021bir,Dong:2021bvy,Ren:2021dsi,Albaladejo:2022sux,Padmanath:2022cvl,Dai:2023mxm}. 
From this perspective, the $T_{cc}$ is not only an outcome of two-body charm dynamics, but also 
a correlated hadronic subsystem that may participate in higher few-body configurations.

Three-body systems containing heavy mesons and a light hadron have already 
been investigated in several contexts, including $D^{(*)}D^{(*)}K$ and 
$D\bar{D}^{(*)}K$ configurations~\cite{Ma:2017ery,Ren:2018pcd,Ren:2024mjh,Zhang:2024yfj,Pan:2025xvq}. 
These studies demonstrate that the addition of a light hadron can substantially reorganize 
the dynamics of a heavy molecular subsystem and may generate bound or resonant structures. 
The pion provides a qualitatively distinct probe. As the lightest hadron and the 
pseudo-Goldstone boson of spontaneously broken chiral symmetry, it governs the 
longest-range component of hadronic interactions and can scatter from both constituents 
of a $D D^*$ molecule. The $\pi T_{cc}$ system therefore offers a natural testing ground for whether an observed hadronic molecule can seed a pion-assisted 
three-body structure.

In this letter, we study the $\pi T_{cc}$ system as a realization of the 
three-body $\pi D D^*$ dynamics. The pion is not treated as a spectator. Specifically, 
its successive interactions with the $D$ and $D^*$ constituents can generate collective 
effects that are absent at the single-scattering level. We formulate the problem within 
the Fixed Center Approximation to the Faddeev equations~\cite{Foldy:1945zz,Deloff:1999gc,Roca:2010tf,Debastiani:2017vhv,Dias:2017miz,Dias:2018iuy,MartinezTorres:2020hus}. 
The $T_{cc}$ is represented as a correlated isoscalar $D D^*$ cluster whose spatial distribution 
is encoded in a form factor, while the elementary $\pi D$ and $\pi D^*$ amplitudes \cite{Guo:2006fu,Guo:2006rp} 
provide the dynamical input for the pion-constituent interactions. The sensitivity 
to the extended and shallow nature of the cluster is assessed by varying the 
form factor parameter, namely the cutoff used to regularize the two-body loop 
function in the unitarized Bethe-Salpeter equation, whose solution determines whether the $D D^*$ interaction generates a molecular state.

We find a broad maximum already at the single-scattering level, near $4100$~MeV, as expected from the two-body amplitude inputs. Conventional FCA shifts it to about $4085$~MeV, and including coherent pion-cluster propagation moves and reshapes it further, producing a near-threshold enhancement around $4055$~MeV. This hierarchy shows that the elementary amplitudes seed the energy dependence, whereas successive three-body mechanisms relocate the strength toward threshold. The corresponding structures in $\mathrm{Re}[T]$ and $\mathrm{Im}[T]$ are compatible with a resonant-like interpretation. The result demonstrates how an experimentally established hadronic molecule can acquire a second role, that is, beyond being the endpoint of two-body dynamics, it can become a building block of a heavier structure shaped by three-body dynamics.

A direct experimental test is provided by the $T_{cc}\pi$ invariant-mass 
distribution, with $R_{\pi T_{cc}}^0\to T_{cc}^+\pi^-\to D^0D^0\pi^+\pi^-$ offering a natural search mode.

\bigskip

\begin{figure*}[t]
\centering
\includegraphics[height=0.3\textheight]{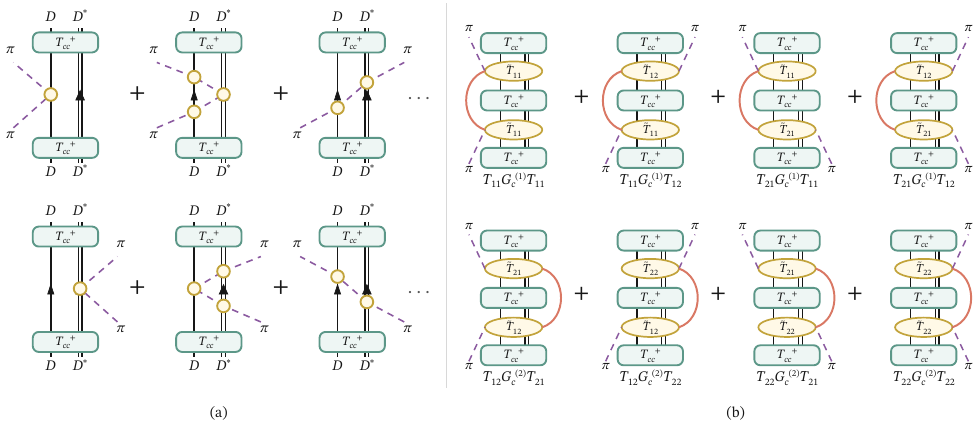}
\caption{Diagrammatic representation of the $\pi T_{cc}$ three-body dynamics: (a) 
Multiple-scattering series defining the conventional Fixed Center 
Approximation (FCA), where the external pion interacts successively 
with the $D$ and $D^*$ constituents of the molecular cluster; (b) Additional 
diagrammatic series arising from the coherent propagation of the intermediate 
$\pi T_{cc}$ state. Their resummation restores the exact elastic-unitarity 
relation near the projectile-cluster threshold while continuously recovering 
the conventional FCA in the limit $G_{c1},G_{c2}\rightarrow0$.}
\label{fig:FCA}
\end{figure*}

{\it Coherent Fixed Center Approximation}. The physical mechanism 
underlying the present calculation is illustrated in Fig.~\ref{fig:FCA}. 
The $T_{cc}(3875)^+$ is treated as a shallow isoscalar $S$-wave $DD^*$ molecular 
cluster with $I(J^P)=0(1^+)$, while the external pion undergoes successive interactions with the $D$ 
and $D^*$ constituents. Taking the relative $\pi T_{cc}$ motion to be in $S$-wave then fixes the three-body channel to $I(J^P)=1(1^-)$. Since the Fixed Center Approximation (FCA) assumes that 
the internal structure of the cluster remains essentially unchanged during the 
scattering process, it provides the appropriate framework for describing the 
low-energy $\pi T_{cc}$ dynamics, where the pion probes the molecular cluster 
without significantly distorting its internal structure.

The conventional multiple-scattering series is represented schematically 
in Fig.~\ref{fig:FCA}(a). In the coherent formulation of the FCA recently developed in 
Ref.~\cite{Ikeno:2025bsx}, the scattering process is organized in terms 
of the partition amplitudes $ T_{ij}$, where $i,j=1,2$ denote the initial 
and final interactions of the pion with the $D$ ($1$) and $D^*$ ($2$) constituents, 
respectively. They satisfy
\begin{align}
 \tilde{T}_{11}
&=
t_1+t_1G_0\tilde T_{21},\nonumber\\
 \tilde{T}_{12}
&=
t_1G_0\tilde T_{22},
\nonumber\\
 \tilde{T}_{21}
&=
t_2G_0\tilde T_{11},\nonumber\\
 \tilde{T}_{22}
&=
t_2+t_2G_0\tilde T_{12},
\end{align}
where $G_0$ describes the propagation of the pion between two successive interactions 
with the constituents of the molecular cluster. Solving these coupled-channel equations and summing the four partitions, 
\begin{equation}
 T=\sum_{i,j} \tilde{T}_{ij},
\end{equation}
one recovers the conventional FCA amplitude, 
\begin{equation}
 T=
\frac{
t_1+t_2+2t_1G_0t_2
}{
1-t_1G_0t_2G_0
}.
\label{eq:FCAconv}
\end{equation}
The elementary amplitudes $t_1$ and $t_2$ correspond to the $\pi D$ and $\pi D^*$ interactions, respectively. 
They enter the three-body calculation through their matrix elements in the $\pi\,(D D^*)_{I=0}$ state. Since the $T_{cc}$ is an isoscalar molecule, by isospin recoupling, they are given by combinations
\begin{equation}
t_i(s_i)=\frac{2}{3}\,t_i^{I=3/2}(s_i)+\frac{1}{3}\,t_i^{I=1/2}(s_i),\qquad i=1,2.    
\end{equation}
The corresponding input amplitudes in the FCA are
\begin{equation}
    t_i(s_i) \rightarrow \tilde{t}_i (s_i) = \frac{M_c}{m_i}\,t_i(s_i)\,,
\end{equation}
where the factors $M_c/m_i$ come from matching the 
relativistic normalization of the pion-constituent amplitudes to that of their single-scattering contributions to the pion-cluster $S$-matrix in the fixed-center limit \cite{Roca:2010tf,Yamagata-Sekihara:2010muv}. This replacement is understood whenever the partition amplitudes are used to construct the physical pion-cluster amplitude. Moreover, the elementary amplitudes are evaluated at the center-of-mass energies $\sqrt{s_i}$ of the corresponding $\pi D$ and $\pi D^*$ subsystems. Within the fixed-center kinematic prescription, these two subenergies are determined from the total three-body invariant mass $\sqrt{s}$ through
\begin{equation}
    s_i = m_\pi^2+m_i^2+\frac{s-m_\pi^2-M_c^2}{2M_c^2}\left(M_c^2+m_i^2-m_j^2\right),\,\, j\neq i\, ,
\end{equation}
where $i=1(2)$ denotes the $D (D^*)$ constituent, $m_1=m_D\,(m_2 = m_{D^*})$, and $j$ labels the other constituent of the cluster. Here, $M_c$ is the $T_{cc}$ mass and $s$ is the squared invariant mass of the full $\pi D D^*$ system.

The amplitudes in definite isospin channels are taken from the unitarized chiral effective 
theory of Refs.~\cite{Guo:2006fu,Guo:2006rp}, employing the parameter set of those works with the 
cutoff fixed to $\Lambda=1000$~MeV. Although more sophisticated coupled-channel descriptions of the 
axial charm sector have recently become available \cite{Brandao:2025cli}, the associated resonances 
lie several hundred MeV above the $\pi T_{cc}$ threshold and therefore have only a minor influence 
on the near-threshold dynamics investigated here.

The essential improvement of the coherent FCA is illustrated in Fig.~\ref{fig:FCA}(b). 
Besides the conventional multiple-scattering mechanism, the intermediate $\pi T_{cc}$ is allowed 
to propagate coherently between successive events. Denoting by $T$ the $2\times 2$ matrix formed by the 
partition amplitudes and by
\begin{equation}
G_c=
\begin{pmatrix}
G_{c1}&0\\
0&G_{c2}
\end{pmatrix},
\end{equation}
the propagator of the pion-cluster state, the full three-body amplitude satisfies the Dyson-type 
equation 
\begin{equation}
T_{\rm tot}=T+TG_cT_{\rm tot},
\end{equation}
whose solution is
\begin{equation}
T_{\rm tot}
=
(I-TG_c)^{-1}T.
\end{equation}
Using the explicit solutions of the partition amplitudes, the physical three-body 
amplitude can be written in the compact form
\begin{equation}
T_{\rm tot}
=
\frac{
\tilde{t}_1+\tilde{t}_2+
\left(
2G_0-G_{c1}-G_{c2}
\right)\tilde{t}_1\tilde{t}_2
}{
1-
\tilde{t}_1G_{c1}-
\tilde{t}_2G_{c2}-
\left(
G_0^2-G_{c1}G_{c2}
\right)\tilde{t}_1\tilde{t}_2
}.
\label{eq:CFCA}
\end{equation}
Eq~\eqref{eq:CFCA} makes explicit that the coherent formulation continuously recovers the 
conventional FCA result, Eq.~\eqref{eq:FCAconv}, in the limit $G_{c1},G_{c2}\rightarrow0$, 
while simultaneously restoring the exact elastic-unitarity relation in the vicinity 
of the projectile-cluster threshold.

Finally, the extended spatial structure of the shallow 
$T_{cc}$ is encoded in the cluster form factor $F(\mathbf q)$, constructed from the $D D^*$ wave function. In the cutoff scheme employed here, the momentum space wave function can be written, up to an overall normalization, 
\begin{align}
\varphi_{DD^*}(\mathbf p)&\equiv\frac{\theta(q_{\max}-|\mathbf p|)}{M_c-\omega_D(p)-\omega_{D^*}(p)},\quad\omega_a(p)=\sqrt{m_a^2+p^2}, 
\end{align}
where $a=D,D^*$, $\theta$ is the Heaviside step function and $q_{\mathrm{max}}$ regularizes the $D D^*$ loop. The elastic cluster form factor is then defined by the normalized overlap
\begin{equation}
    F(\mathbf{q})\equiv\frac{\displaystyle\int\frac{d^3p}{(2\pi)^3}\,\phi^*(\mathbf p)\phi(\mathbf p-\mathbf q)}{\displaystyle\int\frac{d^3p}{(2\pi)^3},|\phi(\mathbf p)|^2}\,,\qquad F(\mathbf 0)=1.
\end{equation}

Since the cluster is an $S$-wave state, the form factor depends only on $q=|\mathbf q|$. It enters the propagator $G_0$, which describes pion propagation between successive interactions with the two constituents
\begin{equation}
    G_0(\sqrt{s})= \int\frac{d^3q}{(2\pi)^3}\,\frac{F(q)}{4\omega_\pi(q)\omega_c(q)}\frac{1}{\sqrt{s}-\omega_\pi(q)-\omega_c(q)+i0}\,,
\end{equation}
where 
\begin{equation}
\omega_\pi(q)=\sqrt{m_\pi^2+q^2},\qquad\omega_c(q)=\sqrt{M_c^2+q^2}.
\end{equation}

The coherent propagators $G_{c1}$ and $G_{c2}$, present in Eq.~\eqref{eq:CFCA}, describe intermediate pion-cluster propagation between scattering events that begin and end at constituent $i$. They are given by
\begin{align}
    G_{ci}(\sqrt{s}) &= \int\frac{d^3q}{(2\pi)^3},\frac{\left[F_c^{(i)}(q)\right]^2}{4\omega_\pi(q)\omega_c(q)}\frac{1}{\sqrt{s}-\omega_\pi(q)-\omega_c(q)+i0},\nonumber \\
    &i=1,2\, ,
\end{align}
with
\begin{equation}
    F^{(i)}_c(\alpha_i q)=F(\alpha_i\,q),\qquad\alpha_i=\frac{m_j}{m_i+m_j},\, j\neq i. 
\end{equation}
The factor $\alpha_i$ accounts for the fraction of the transferred momentum carried by constituent $i$ inside the cluster \cite{Yamagata-Sekihara:2010kpd,Encarnacion:2026zas}, whereas the square of $F_c^{(i)}$ arises from the cluster overlap at the two ends of the coherent propagation. 
We use $q_{\max}=900$ MeV for the central result, and define the uncertainty bands as the point-by-point envelopes obtained with $q_{\max}=700$, $900$, and $1100$ MeV.

% ---------------------------------------------------------------------
% ---------------------------------------------------------------------
% -------------------------------------------------------------------

\bigskip

{\it Results and discussion.}--- We now present the numerical results  for the elastic scattering of a pion from the $T_{cc}$ cluster. The 
amplitude is evaluated as a function of the center-of-mass $\sqrt{s}$ 
energy of the external pion-cluster system. Throughout the calculation 
we work in the isospin basis and we use isospin-averaged masses for 
the pion and charmed mesons, consistently with the two-body amplitudes 
entering the FCA equations. 
The $\pi\,T_{cc}$ threshold is indicated 
in the figures to make the near-threshold nature of the resulting 
structure explicit. 

\begin{figure}[!tp]
    \centering
    \subfigure[]{
        \centering
        \includegraphics[width=0.45\textwidth]{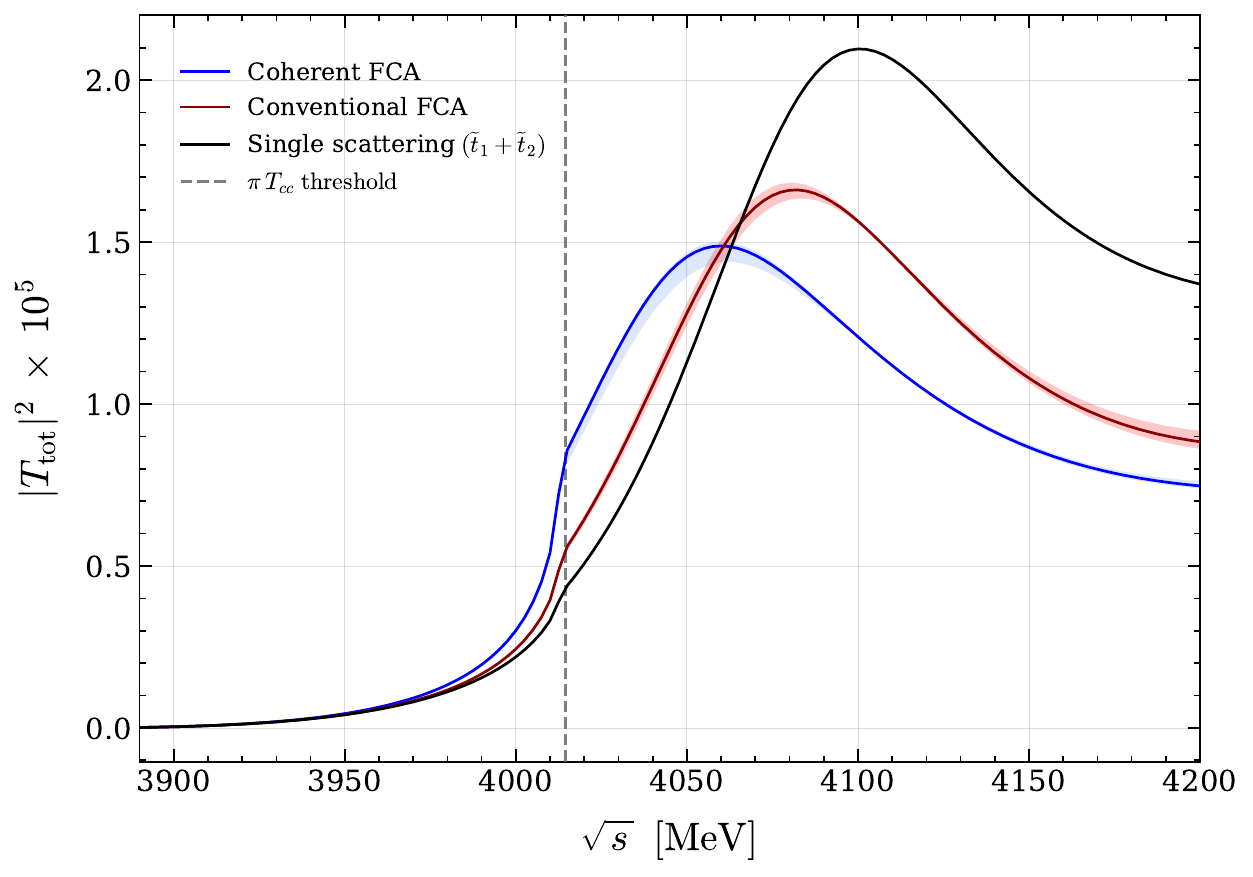}
%        \caption{}
        \label{fig:T_a}
    }
    \subfigure[]{
        \centering
        \includegraphics[width=0.45\textwidth]{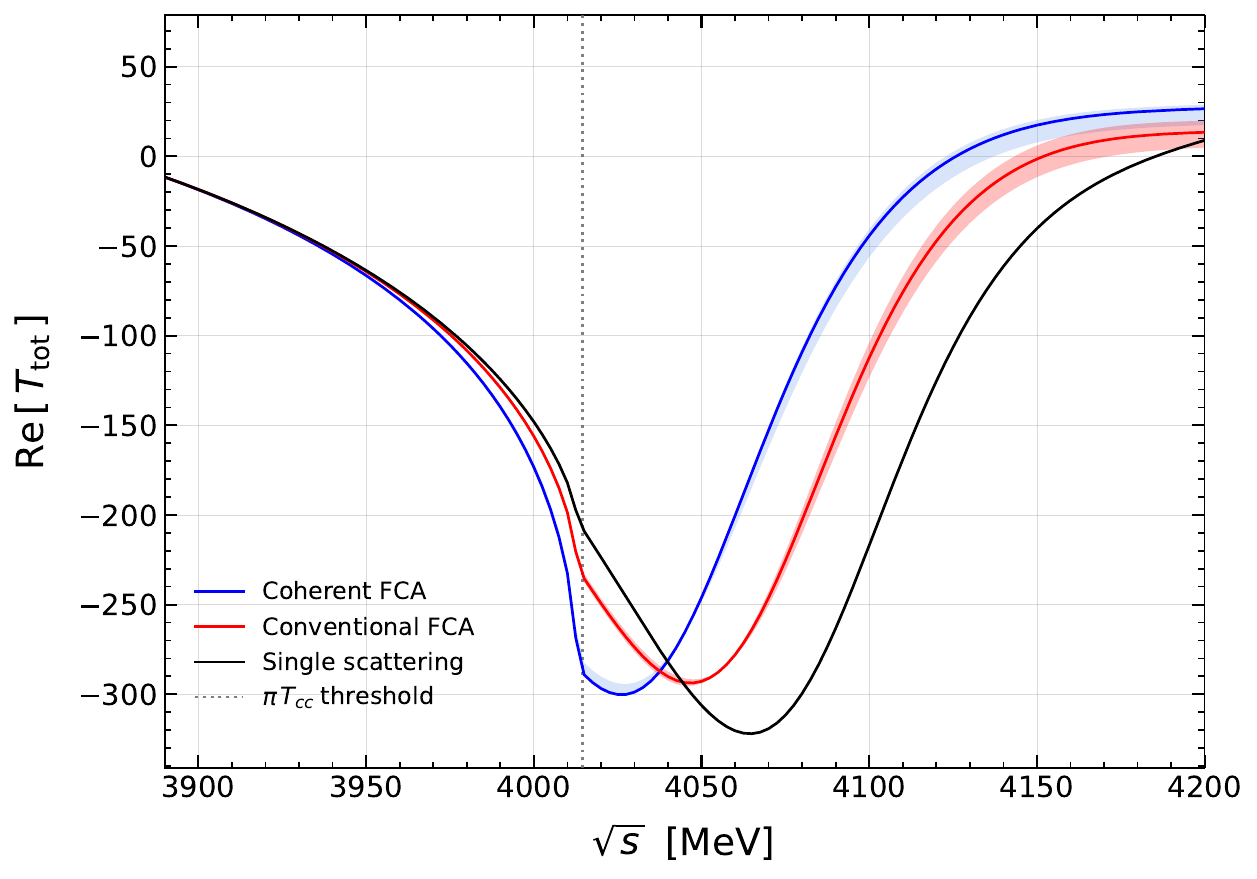}
%        \caption{}
        \label{fig:T_b}
    }
    \subfigure[]{
        \centering
        \includegraphics[width=0.45\textwidth]{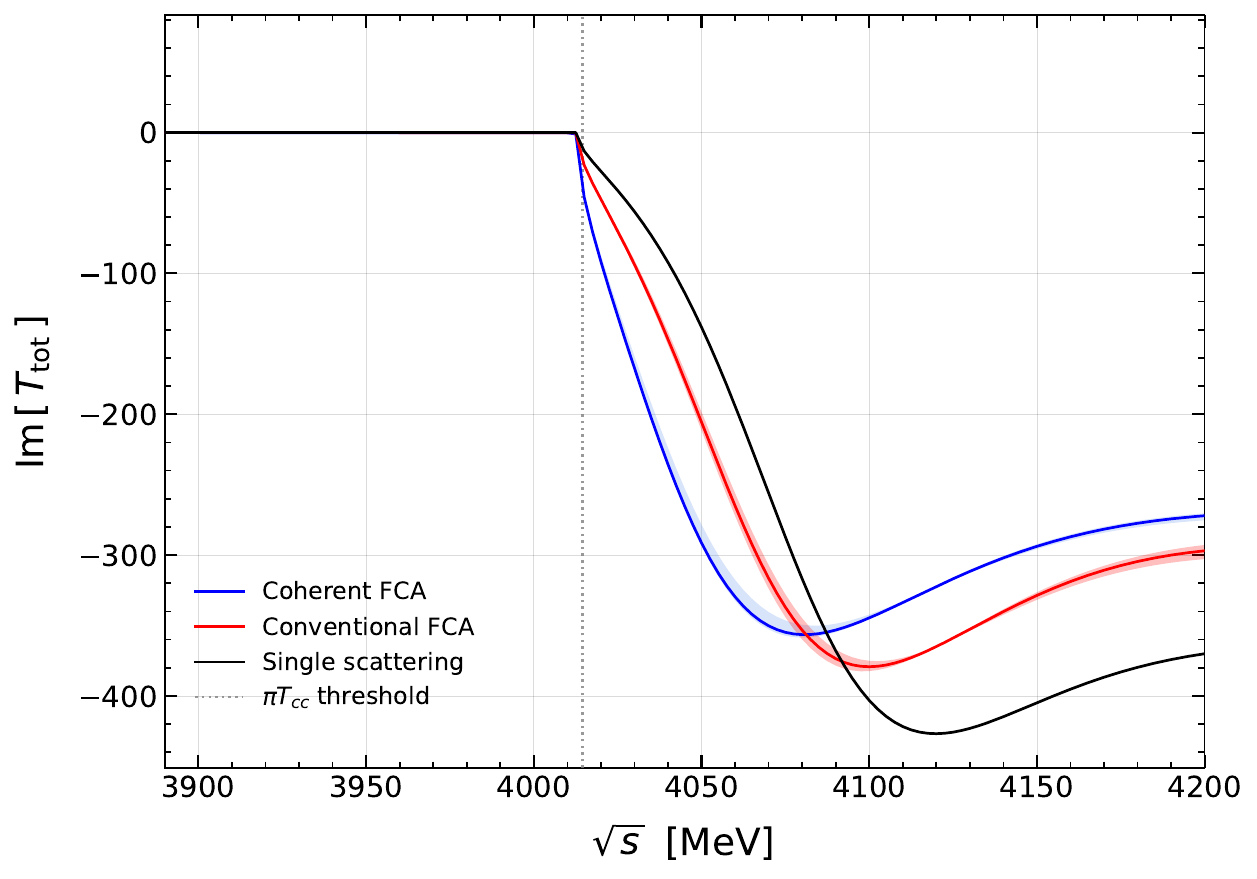}
%        \caption{}
        \label{fig:T_c}
    }
\caption{
Panel \ref{fig:T_a} displays the squared $\pi T_{cc}$ amplitude in the $I(J^P)=1(1^-)$ channel for three successive 
approximations. The black line denotes the single scattering $\tilde{t}_1+\tilde{t}_2$ (i.e. taking $G_0=G_{c1}=G_{c2}=0$ in 
Eq.~\eqref{eq:CFCA}); the red line is conventional FCA, obtained by setting $G_{c1}=G_{c2}=0$; and the blue line is the full coherent result for 
$q_{\max}=900$~MeV. The blue and red bands are the point-by-point envelopes of the results obtained with $q_{\max}=700$--$1100$~MeV; and the 
vertical line marks the $\pi T_{cc}$ threshold. Panels \ref{fig:T_b} and \ref{fig:T_c} show respectively the real and imaginary parts of the $\pi T_{cc}$  amplitude, using the same color scheme as in Panel \ref{fig:T_a}. }
\label{fig:T_abc}
\end{figure}

The bands shown in Fig.~\ref{fig:T_abc} quantify the sensitivity of the three-body amplitude to the \(T_{cc}\) molecular form factor. They are constructed as the point-by-point envelopes of the results obtained with \(q_{\max}=700\), \(900\), and \(1100\) MeV, with \(q_{\max}=900\) MeV defining the central result, while the cutoff entering the elementary \(\pi D\) and \(\pi D^*\) amplitudes is kept fixed at \(\Lambda=1000\) MeV. This prescription tests the stability of the predicted near-threshold enhancement against variations of the cluster form factor.

Figure~\ref{fig:T_a} resolves the dynamical origin of the line shape. For the 
central parameter set, single scattering already produces a broad maximum near 
$4100$~MeV. Resumming the conventional FCA series through $G_0$ shifts the maximum 
to about $4085$~MeV. The coherent propagators $G_{c1}$ and $G_{c2}$ then move the 
strength further toward threshold and reshape it into a maximum near $4060$~MeV. 
Varying $q_{\max}$ leaves this qualitative progression intact and places the full 
maximum in the approximate range $4050$--$4075$~MeV. Thus, the final enhancement is 
not generated from a featureless single-scattering background: the elementary 
$\pi D$ and $\pi D^*$ amplitudes seed the broad structure, while multiple scattering 
and coherent pion--cluster propagation determine its near-threshold location and 
line shape.

The location of the enhancement also provides a useful check of the momentum scale involved in the pion-cluster motion. Since the peak appears about $45$~MeV above the $\pi\,T_{cc}$ threshold, the corresponding relative momentum in the $\pi\,T_{cc}$ center-of-mass frame is $|\vek_{\pi T_{cc}}|\simeq 110$~MeV. Since $M_{T_{cc}}\gg m_\pi$, the reduced mass of the system is close to the pion mass, and this relative momentum can be interpreted as the momentum scale with which the pion probes the cluster. It corresponds to a length scale $1/|\vek_{\pi T_{cc}}|\sim 2$~fm, which is of the order of typical hadronic molecular distances \cite{Guo:2017jvc}. Thus, the pion carries a wavelength short enough to be sensitive to the extended $DD^*$ structure of the $T_{cc}$, while still probing the system at a hadronic scale. This provides a consistency check for a three-body treatment in which the pion interacts with the individual constituents of the molecular cluster. Recall that in the FCA calculation, the $DD^*$ correlation is retained through the cluster form factor, and the residual sensitivity to possible distortions of the cluster is included in the uncertainty band discussed above.

The behavior of the complex amplitude further supports the interpretation of the enhancement as a resonant-like structure in the $I(J^P)=1(1^-)$ channel. 
As shown in Figs.~\ref{fig:T_b} and ~\ref{fig:T_c}, the peak in $|T|^2$ is accompanied by a rapid variation of $\text{Re}[T]$ and by a corresponding 
structure in $\text{Im}[T]$ in the same region. This is the characteristic 
pattern expected from a resonant amplitude near threshold. 
The comparison between the conventional FCA and the coherent propagation is particularly instructive, and provides deeper insight into the underlying dynamics. In the coherent amplitude, both the real and imaginary parts are shifted toward the threshold, with the imaginary part being slightly suppressed in magnitude. These modifications originate from the interference between the coherent contributions mediated by $G_{c1}$ and $G_{c2}$ and the conventional $G_0$ contribution in Eq.~\eqref{eq:CFCA}. Since both the $t_1$ and $t_2$ two-body amplitudes exhibit resonant behavior, the observed  effect cannot be interpreted as a simple sum of individual contributions. Rather, it emerges from the coherent interference between these two resonant channels, with the lower poles associated with the resonant structures being the most relevant.

Together with the comparison to the single-scattering result, it indicates that the 
signal is associated with the three-body dynamics of the $\pi\,DD^*$ 
system rather than with a trivial kinematical effect at threshold. To characterize the near-threshold interaction quantitatively, we follow the effective range procedure of Ref.~\cite{Agatao:2025ckp}, adapted to the meson-meson normalization employed here. The quantum-mechanical scattering amplitude is related to the three-body amplitude through
\begin{equation}
\left(f_{\pi T_{cc}}^{\rm QM}\right)^{-1}
=
-8\pi\sqrt{s}\,T_{\rm tot}^{-1}
\simeq
-\frac{1}{a}
+\frac{1}{2}r_0 k^2
-ik ,
\label{eq:ERE}
\end{equation}
where
\begin{equation}
k=
\frac{\lambda^{1/2}(s,m_\pi^2,M_c^2)}
     {2\sqrt{s}} \, ,
\label{eq:relative-momentum}
\end{equation}
is the relative pion-cluster momentum and $\lambda$ denotes de Källén function. The scattering parameters are then obtained as 
\begin{equation}
\begin{aligned}
a &=
\left.
\frac{T_{\rm tot}}
     {8\pi\sqrt{s}}
\right|_{\rm th},
\\[2pt]
r_0 &=
\frac{1}{\mu}
\left[
\frac{\partial}{\partial\sqrt{s}}
\left(
-8\pi\sqrt{s}\,T_{\rm tot}^{-1}
+ik
\right)
\right]_{\rm th},
\end{aligned}
\label{eq:scattering-parameters}
\end{equation}
with
\begin{equation}
\mu=\frac{m_\pi M_c}{m_\pi+M_c},
\qquad
\sqrt{s_{\rm th}}=m_\pi+M_c .
\label{eq:threshold-kinematics}
\end{equation}
Near threshold, the energy and relative momentum are related by $\sqrt{s}=m_\pi+M_c+k^2/(2\mu)$. The subscript ``${\rm th}$'' denotes evaluation at the $\pi T_{cc}$ threshold, and our convention corresponds to $f_{\pi T_{cc}}^{\rm QM}(0)=-a$. For the central value $q_{\max}=900~{\rm MeV}$, we obtain
\begin{equation}
a=-0.557~{\rm fm},
\qquad
r_0=(-8.56-i\,6.68)~{\rm fm}.
\label{eq:scattering-parameter-values}
\end{equation}
The large negative effective range indicates an interaction characteristic of pion-exchange dynamics in the $\pi T_{cc}$ system, supporting the interpretation of the observed enhancement at $M \simeq 4055$~MeV as a resonant-like three-body structure arising from coherent multiple scattering of the pion from the $D$ and $D^*$ components of the $T_{cc}$ cluster, not from a simple two-body interaction.

It should be mentioned that we have checked the results using distinct approaches for the elementary amplitudes. For example, we have employed the amplitude for the $\pi D^{\ast}$ system from Ref.~\cite{Brandao:2025cli}, and the findings did not exhibit significant deviations.
This is primarily because the threshold for the $\pi T_{cc}$ system was set at $\sim 4013$~MeV, which corresponds to approximately $2100$~MeV in the invariant mass scale of the two-body subsystems. More importantly, this threshold lies about $300$~MeV away from the pole positions associated to the lightest axial charmed resonances. Such a substantial energy gap suppresses any possible dynamical enhancement or interference effects that could otherwise arise from the proximity of this charmed meson resonance. Consequently, the limited sensitivity to the higher poles ensures that our amplitude parametrization remains almost unaffected by its presence, thereby justifying the observed similarity in our results.

Let us now contextualize the experimental environment in which such a structure could be observed. The most direct experimental signature of such a state would be 
an enhancement in the $T_{cc}$ invariant-mass distribution. Since the 
$T^+_{cc}$ has been observed through the $D^0D^0\pi^+$ final state, 
a natural search channel is 
$R^0_{\pi T_{cc}} \to T^+_{cc}\pi^-\to D^0D^0\pi^+\pi^-$. In an experimental 
analysis, the near-threshold structure could be studied either in the 
invariant mass $M(D^0D^0\pi^+\pi^-)$ or, more conveniently, in the energy 
released with respect to the $\pi\,T_{cc}$ threshold, 
$T=M(D^0D^0\pi^+\pi^-) - M(D^0D^0\pi^+) - m_{\pi^-}$.

{\it Concluding remarks.} We have investigated the pion scattering off the $T_{cc}$, treated as a shallow isoscalar $D D^*$ molecular cluster, within the coherent Fixed Center Approximation. The comparison between each approximation clarifies the origin of the predicted line shape. At the single-scattering level, the energy dependence inherited from the elementary $\pi D$ and $\pi D^*$ amplitudes produces a broad maximum near $4.10$ GeV. Conventional three-body rescattering through $G_0$ and the subsequent coherent pion-cluster propagation lead to distinct resummed amplitudes, progressively redistributing the strength toward the $\pi T_{cc}$ threshold and producing an $I(J^P)=1(1^-)$ enhancement near $4.06$ GeV. This sequence demonstrates that genuine three-body mechanisms play a decisive role in determining the position and shape of the final enhancement. The qualitative behavior remains stable under variations of the molecular form factor cutoff and under changes in the elementary two-body inputs.

The broader implication is that the pion does not act merely as a passive probe of a preexisting $T_{cc}$ molecule. Through its repeated interactions with the $D$ and $D^*$ constituents, it actively reorganizes the dynamics of the molecular subsystem. An experimental observation of the predicted enhancement in the $\pi T_{cc}$ invariant-mass distribution, particularly in the $D^0 D^0 \pi^+ \pi^-$ channel, would therefore provide evidence that an established hadronic molecule can serve as a dynamical building block of a higher few-body configuration. The nontrivial energy dependence found here also motivates extending the coherent FCA to complex energies. Such a development would make it possible to map the associated singularity structure and investigate whether the enhancement is connected to a nearby three-body pole. The $\pi T_{cc}$ system thus provides both an experimentally accessible manifestation of few-body hadronic dynamics and a concrete starting point for a more general analytic formulation of the coherent FCA.

{\it Acknowledgments.}
This work was partly supported by the Brazilian agencies CNPq  (Conselho Nacional de Desenvolvimento Cient\'ifico e Tecnol\'ogico) (L.M.A.: Grants No. 400215/2022-5, 308299/2023-0, 402942/2024-8) and CNPq/FAPERJ under the Project INCT-F\'isica Nuclear e Aplica\c{c}\~oes (Contract No. 408419/2024-5).

\bibliography{refs}

%apsrev4-2.bst 2019-01-14 (MD) hand-edited version of apsrev4-1.bst
%Control: key (0)
%Control: author (8) initials jnrlst
%Control: editor formatted (1) identically to author
%Control: production of article title (0) allowed
%Control: page (0) single
%Control: year (1) truncated
%Control: production of eprint (0) enabled
\begin{thebibliography}{31}%
\makeatletter
\providecommand \@ifxundefined [1]{%
 \@ifx{#1\undefined}
}%
\providecommand \@ifnum [1]{%
 \ifnum #1\expandafter \@firstoftwo
 \else \expandafter \@secondoftwo
 \fi
}%
\providecommand \@ifx [1]{%
 \ifx #1\expandafter \@firstoftwo
 \else \expandafter \@secondoftwo
 \fi
}%
\providecommand \natexlab [1]{#1}%
\providecommand \enquote  [1]{``#1''}%
\providecommand \bibnamefont  [1]{#1}%
\providecommand \bibfnamefont [1]{#1}%
\providecommand \citenamefont [1]{#1}%
\providecommand \href@noop [0]{\@secondoftwo}%
\providecommand \href [0]{\begingroup \@sanitize@url \@href}%
\providecommand \@href[1]{\@@startlink{#1}\@@href}%
\providecommand \@@href[1]{\endgroup#1\@@endlink}%
\providecommand \@sanitize@url [0]{\catcode `\\12\catcode `\$12\catcode
  `\&12\catcode `\#12\catcode `\^12\catcode `\_12\catcode `\%12\relax}%
\providecommand \@@startlink[1]{}%
\providecommand \@@endlink[0]{}%
\providecommand \url  [0]{\begingroup\@sanitize@url \@url }%
\providecommand \@url [1]{\endgroup\@href {#1}{\urlprefix }}%
\providecommand \urlprefix  [0]{URL }%
\providecommand \Eprint [0]{\href }%
\providecommand \doibase [0]{https://doi.org/}%
\providecommand \selectlanguage [0]{\@gobble}%
\providecommand \bibinfo  [0]{\@secondoftwo}%
\providecommand \bibfield  [0]{\@secondoftwo}%
\providecommand \translation [1]{[#1]}%
\providecommand \BibitemOpen [0]{}%
\providecommand \bibitemStop [0]{}%
\providecommand \bibitemNoStop [0]{.\EOS\space}%
\providecommand \EOS [0]{\spacefactor3000\relax}%
\providecommand \BibitemShut  [1]{\csname bibitem#1\endcsname}%
\let\auto@bib@innerbib\@empty
%</preamble>
\bibitem [{\citenamefont {Guo}\ \emph {et~al.}(2018)\citenamefont {Guo},
  \citenamefont {Hanhart}, \citenamefont {Mei{\ss}ner}, \citenamefont {Wang},
  \citenamefont {Zhao},\ and\ \citenamefont {Zou}}]{Guo:2017jvc}%
  \BibitemOpen
  \bibfield  {author} {\bibinfo {author} {\bibfnamefont {F.-K.}\ \bibnamefont
  {Guo}}, \bibinfo {author} {\bibfnamefont {C.}~\bibnamefont {Hanhart}},
  \bibinfo {author} {\bibfnamefont {U.-G.}\ \bibnamefont {Mei{\ss}ner}},
  \bibinfo {author} {\bibfnamefont {Q.}~\bibnamefont {Wang}}, \bibinfo {author}
  {\bibfnamefont {Q.}~\bibnamefont {Zhao}},\ and\ \bibinfo {author}
  {\bibfnamefont {B.-S.}\ \bibnamefont {Zou}},\ }\bibfield  {title} {\bibinfo
  {title} {{Hadronic molecules}},\ }\href
  {https://doi.org/10.1103/RevModPhys.90.015004} {\bibfield  {journal}
  {\bibinfo  {journal} {Rev. Mod. Phys.}\ }\textbf {\bibinfo {volume} {90}},\
  \bibinfo {pages} {015004} (\bibinfo {year} {2018})},\ \bibinfo {note}
  {[Erratum: Rev.Mod.Phys. 94, 029901 (2022)]},\ \Eprint
  {https://arxiv.org/abs/1705.00141} {arXiv:1705.00141 [hep-ph]} \BibitemShut
  {NoStop}%
\bibitem [{\citenamefont {Liu}\ \emph {et~al.}(2025)\citenamefont {Liu},
  \citenamefont {Pan}, \citenamefont {Liu}, \citenamefont {Wu}, \citenamefont
  {Lu},\ and\ \citenamefont {Geng}}]{Liu:2024uxn}%
  \BibitemOpen
  \bibfield  {author} {\bibinfo {author} {\bibfnamefont {M.-Z.}\ \bibnamefont
  {Liu}}, \bibinfo {author} {\bibfnamefont {Y.-W.}\ \bibnamefont {Pan}},
  \bibinfo {author} {\bibfnamefont {Z.-W.}\ \bibnamefont {Liu}}, \bibinfo
  {author} {\bibfnamefont {T.-W.}\ \bibnamefont {Wu}}, \bibinfo {author}
  {\bibfnamefont {J.-X.}\ \bibnamefont {Lu}},\ and\ \bibinfo {author}
  {\bibfnamefont {L.-S.}\ \bibnamefont {Geng}},\ }\bibfield  {title} {\bibinfo
  {title} {{Three ways to decipher the nature of exotic hadrons: Multiplets,
  three-body hadronic molecules, and correlation functions}},\ }\href
  {https://doi.org/10.1016/j.physrep.2024.12.001} {\bibfield  {journal}
  {\bibinfo  {journal} {Phys. Rept.}\ }\textbf {\bibinfo {volume} {1108}},\
  \bibinfo {pages} {1} (\bibinfo {year} {2025})},\ \Eprint
  {https://arxiv.org/abs/2404.06399} {arXiv:2404.06399 [hep-ph]} \BibitemShut
  {NoStop}%
\bibitem [{\citenamefont {Wang}\ \emph {et~al.}(2026)\citenamefont {Wang},
  \citenamefont {Liu},\ and\ \citenamefont {Gao}}]{Wang:2025dur}%
  \BibitemOpen
  \bibfield  {author} {\bibinfo {author} {\bibfnamefont {X.}~\bibnamefont
  {Wang}}, \bibinfo {author} {\bibfnamefont {X.}~\bibnamefont {Liu}},\ and\
  \bibinfo {author} {\bibfnamefont {Y.}~\bibnamefont {Gao}},\ }\bibfield
  {title} {\bibinfo {title} {{Colloquium: Hadron production in open-charm meson
  pairs at e+e- colliders}},\ }\href {https://doi.org/10.1103/2mrp-chly}
  {\bibfield  {journal} {\bibinfo  {journal} {Rev. Mod. Phys.}\ }\textbf
  {\bibinfo {volume} {98}},\ \bibinfo {pages} {021001} (\bibinfo {year}
  {2026})},\ \Eprint {https://arxiv.org/abs/2502.15117} {arXiv:2502.15117
  [hep-ex]} \BibitemShut {NoStop}%
\bibitem [{\citenamefont {Aaij}\ \emph
  {et~al.}(2022{\natexlab{a}})\citenamefont {Aaij} \emph
  {et~al.}}]{LHCb:2021auc}%
  \BibitemOpen
  \bibfield  {author} {\bibinfo {author} {\bibfnamefont {R.}~\bibnamefont
  {Aaij}} \emph {et~al.} (\bibinfo {collaboration} {LHCb}),\ }\bibfield
  {title} {\bibinfo {title} {{Study of the doubly charmed tetraquark
  $T_{cc}^{+}$}},\ }\href {https://doi.org/10.1038/s41467-022-30206-w}
  {\bibfield  {journal} {\bibinfo  {journal} {Nature Commun.}\ }\textbf
  {\bibinfo {volume} {13}},\ \bibinfo {pages} {3351} (\bibinfo {year}
  {2022}{\natexlab{a}})},\ \Eprint {https://arxiv.org/abs/2109.01056}
  {arXiv:2109.01056 [hep-ex]} \BibitemShut {NoStop}%
\bibitem [{\citenamefont {Aaij}\ \emph
  {et~al.}(2022{\natexlab{b}})\citenamefont {Aaij} \emph
  {et~al.}}]{LHCb:2021vvq}%
  \BibitemOpen
  \bibfield  {author} {\bibinfo {author} {\bibfnamefont {R.}~\bibnamefont
  {Aaij}} \emph {et~al.} (\bibinfo {collaboration} {LHCb}),\ }\bibfield
  {title} {\bibinfo {title} {{Observation of an exotic narrow doubly charmed
  tetraquark}},\ }\href {https://doi.org/10.1038/s41567-022-01614-y} {\bibfield
   {journal} {\bibinfo  {journal} {Nature Phys.}\ }\textbf {\bibinfo {volume}
  {18}},\ \bibinfo {pages} {751} (\bibinfo {year} {2022}{\natexlab{b}})},\
  \Eprint {https://arxiv.org/abs/2109.01038} {arXiv:2109.01038 [hep-ex]}
  \BibitemShut {NoStop}%
\bibitem [{\citenamefont {Ling}\ \emph {et~al.}(2022)\citenamefont {Ling},
  \citenamefont {Liu}, \citenamefont {Geng}, \citenamefont {Wang},\ and\
  \citenamefont {Xie}}]{Ling:2021bir}%
  \BibitemOpen
  \bibfield  {author} {\bibinfo {author} {\bibfnamefont {X.~Z.}\ \bibnamefont
  {Ling}}, \bibinfo {author} {\bibfnamefont {M.~Z.}\ \bibnamefont {Liu}},
  \bibinfo {author} {\bibfnamefont {L.~S.}\ \bibnamefont {Geng}}, \bibinfo
  {author} {\bibfnamefont {E.}~\bibnamefont {Wang}},\ and\ \bibinfo {author}
  {\bibfnamefont {J.~J.}\ \bibnamefont {Xie}},\ }\bibfield  {title} {\bibinfo
  {title} {{Can we understand the decay width of the $T_{cc}^+$ state?}},\
  }\href {https://doi.org/10.1016/j.physletb.2022.136897} {\bibfield  {journal}
  {\bibinfo  {journal} {Phys. Lett. B}\ }\textbf {\bibinfo {volume} {826}},\
  \bibinfo {pages} {136897} (\bibinfo {year} {2022})},\ \Eprint
  {https://arxiv.org/abs/2108.00947} {arXiv:2108.00947 [hep-ph]} \BibitemShut
  {NoStop}%
\bibitem [{\citenamefont {Dong}\ \emph {et~al.}(2021)\citenamefont {Dong},
  \citenamefont {Guo},\ and\ \citenamefont {Zou}}]{Dong:2021bvy}%
  \BibitemOpen
  \bibfield  {author} {\bibinfo {author} {\bibfnamefont {X.-K.}\ \bibnamefont
  {Dong}}, \bibinfo {author} {\bibfnamefont {F.-K.}\ \bibnamefont {Guo}},\ and\
  \bibinfo {author} {\bibfnamefont {B.-S.}\ \bibnamefont {Zou}},\ }\bibfield
  {title} {\bibinfo {title} {{A survey of heavy-heavy hadronic molecules}},\
  }\href {https://doi.org/10.1088/1572-9494/ac27a2} {\bibfield  {journal}
  {\bibinfo  {journal} {Commun. Theor. Phys.}\ }\textbf {\bibinfo {volume}
  {73}},\ \bibinfo {pages} {125201} (\bibinfo {year} {2021})},\ \Eprint
  {https://arxiv.org/abs/2108.02673} {arXiv:2108.02673 [hep-ph]} \BibitemShut
  {NoStop}%
\bibitem [{\citenamefont {Ren}\ \emph {et~al.}(2022)\citenamefont {Ren},
  \citenamefont {Wu},\ and\ \citenamefont {Zhu}}]{Ren:2021dsi}%
  \BibitemOpen
  \bibfield  {author} {\bibinfo {author} {\bibfnamefont {H.}~\bibnamefont
  {Ren}}, \bibinfo {author} {\bibfnamefont {F.}~\bibnamefont {Wu}},\ and\
  \bibinfo {author} {\bibfnamefont {R.}~\bibnamefont {Zhu}},\ }\bibfield
  {title} {\bibinfo {title} {{Hadronic Molecule Interpretation of Tcc+ and Its
  Beauty Partners}},\ }\href {https://doi.org/10.1155/2022/9103031} {\bibfield
  {journal} {\bibinfo  {journal} {Adv. High Energy Phys.}\ }\textbf {\bibinfo
  {volume} {2022}},\ \bibinfo {pages} {9103031} (\bibinfo {year} {2022})},\
  \Eprint {https://arxiv.org/abs/2109.02531} {arXiv:2109.02531 [hep-ph]}
  \BibitemShut {NoStop}%
\bibitem [{\citenamefont {Albaladejo}\ and\ \citenamefont
  {Nieves}(2022)}]{Albaladejo:2022sux}%
  \BibitemOpen
  \bibfield  {author} {\bibinfo {author} {\bibfnamefont {M.}~\bibnamefont
  {Albaladejo}}\ and\ \bibinfo {author} {\bibfnamefont {J.}~\bibnamefont
  {Nieves}},\ }\bibfield  {title} {\bibinfo {title} {{Compositeness of S-wave
  weakly-bound states from next-to-leading order Weinberg{\textquoteright}s
  relations}},\ }\href {https://doi.org/10.1140/epjc/s10052-022-10695-1}
  {\bibfield  {journal} {\bibinfo  {journal} {Eur. Phys. J. C}\ }\textbf
  {\bibinfo {volume} {82}},\ \bibinfo {pages} {724} (\bibinfo {year} {2022})},\
  \Eprint {https://arxiv.org/abs/2203.04864} {arXiv:2203.04864 [hep-ph]}
  \BibitemShut {NoStop}%
\bibitem [{\citenamefont {Padmanath}\ and\ \citenamefont
  {Prelovsek}(2022)}]{Padmanath:2022cvl}%
  \BibitemOpen
  \bibfield  {author} {\bibinfo {author} {\bibfnamefont {M.}~\bibnamefont
  {Padmanath}}\ and\ \bibinfo {author} {\bibfnamefont {S.}~\bibnamefont
  {Prelovsek}},\ }\bibfield  {title} {\bibinfo {title} {{Signature of a Doubly
  Charm Tetraquark Pole in DD* Scattering on the Lattice}},\ }\href
  {https://doi.org/10.1103/PhysRevLett.129.032002} {\bibfield  {journal}
  {\bibinfo  {journal} {Phys. Rev. Lett.}\ }\textbf {\bibinfo {volume} {129}},\
  \bibinfo {pages} {032002} (\bibinfo {year} {2022})},\ \Eprint
  {https://arxiv.org/abs/2202.10110} {arXiv:2202.10110 [hep-lat]} \BibitemShut
  {NoStop}%
\bibitem [{\citenamefont {Dai}\ \emph {et~al.}(2023)\citenamefont {Dai},
  \citenamefont {Fleming}, \citenamefont {Hodges},\ and\ \citenamefont
  {Mehen}}]{Dai:2023mxm}%
  \BibitemOpen
  \bibfield  {author} {\bibinfo {author} {\bibfnamefont {L.}~\bibnamefont
  {Dai}}, \bibinfo {author} {\bibfnamefont {S.}~\bibnamefont {Fleming}},
  \bibinfo {author} {\bibfnamefont {R.}~\bibnamefont {Hodges}},\ and\ \bibinfo
  {author} {\bibfnamefont {T.}~\bibnamefont {Mehen}},\ }\bibfield  {title}
  {\bibinfo {title} {{Strong decays of Tcc+ at NLO in an effective field
  theory}},\ }\href {https://doi.org/10.1103/PhysRevD.107.076001} {\bibfield
  {journal} {\bibinfo  {journal} {Phys. Rev. D}\ }\textbf {\bibinfo {volume}
  {107}},\ \bibinfo {pages} {076001} (\bibinfo {year} {2023})},\ \Eprint
  {https://arxiv.org/abs/2301.11950} {arXiv:2301.11950 [hep-ph]} \BibitemShut
  {NoStop}%
\bibitem [{\citenamefont {Ma}\ \emph {et~al.}(2019)\citenamefont {Ma},
  \citenamefont {Wang},\ and\ \citenamefont {Mei{\ss}ner}}]{Ma:2017ery}%
  \BibitemOpen
  \bibfield  {author} {\bibinfo {author} {\bibfnamefont {L.}~\bibnamefont
  {Ma}}, \bibinfo {author} {\bibfnamefont {Q.}~\bibnamefont {Wang}},\ and\
  \bibinfo {author} {\bibfnamefont {U.-G.}\ \bibnamefont {Mei{\ss}ner}},\
  }\bibfield  {title} {\bibinfo {title} {{Double heavy tri-hadron bound state
  via delocalized $\pi$ bond}},\ }\href
  {https://doi.org/10.1088/1674-1137/43/1/014102} {\bibfield  {journal}
  {\bibinfo  {journal} {Chin. Phys. C}\ }\textbf {\bibinfo {volume} {43}},\
  \bibinfo {pages} {014102} (\bibinfo {year} {2019})},\ \Eprint
  {https://arxiv.org/abs/1711.06143} {arXiv:1711.06143 [hep-ph]} \BibitemShut
  {NoStop}%
\bibitem [{\citenamefont {Ren}\ \emph {et~al.}(2018)\citenamefont {Ren},
  \citenamefont {Malabarba}, \citenamefont {Geng}, \citenamefont
  {Khemchandani},\ and\ \citenamefont {Mart{\'\i}nez~Torres}}]{Ren:2018pcd}%
  \BibitemOpen
  \bibfield  {author} {\bibinfo {author} {\bibfnamefont {X.-L.}\ \bibnamefont
  {Ren}}, \bibinfo {author} {\bibfnamefont {B.~B.}\ \bibnamefont {Malabarba}},
  \bibinfo {author} {\bibfnamefont {L.-S.}\ \bibnamefont {Geng}}, \bibinfo
  {author} {\bibfnamefont {K.~P.}\ \bibnamefont {Khemchandani}},\ and\ \bibinfo
  {author} {\bibfnamefont {A.}~\bibnamefont {Mart{\'\i}nez~Torres}},\
  }\bibfield  {title} {\bibinfo {title} {{$K^*$ mesons with hidden charm
  arising from $KX(3872)$ and $KZ_c(3900)$ dynamics}},\ }\href
  {https://doi.org/10.1016/j.physletb.2018.08.034} {\bibfield  {journal}
  {\bibinfo  {journal} {Phys. Lett. B}\ }\textbf {\bibinfo {volume} {785}},\
  \bibinfo {pages} {112} (\bibinfo {year} {2018})},\ \Eprint
  {https://arxiv.org/abs/1805.08330} {arXiv:1805.08330 [hep-ph]} \BibitemShut
  {NoStop}%
\bibitem [{\citenamefont {Ren}\ \emph {et~al.}(2024)\citenamefont {Ren},
  \citenamefont {Khemchandani},\ and\ \citenamefont
  {Mart{\'\i}nez~Torres}}]{Ren:2024mjh}%
  \BibitemOpen
  \bibfield  {author} {\bibinfo {author} {\bibfnamefont {X.-L.}\ \bibnamefont
  {Ren}}, \bibinfo {author} {\bibfnamefont {K.~P.}\ \bibnamefont
  {Khemchandani}},\ and\ \bibinfo {author} {\bibfnamefont {A.}~\bibnamefont
  {Mart{\'\i}nez~Torres}},\ }\bibfield  {title} {\bibinfo {title} {{Heavy $K^*$
  mesons with open charm from $KD^{(*)}D^*$ interactions}},\ }\href
  {https://doi.org/10.1140/epjc/s10052-024-13683-9} {\bibfield  {journal}
  {\bibinfo  {journal} {Eur. Phys. J. C}\ }\textbf {\bibinfo {volume} {84}},\
  \bibinfo {pages} {1297} (\bibinfo {year} {2024})},\ \Eprint
  {https://arxiv.org/abs/2409.16281} {arXiv:2409.16281 [hep-ph]} \BibitemShut
  {NoStop}%
\bibitem [{\citenamefont {Zhang}\ \emph {et~al.}(2025)\citenamefont {Zhang},
  \citenamefont {Hu}, \citenamefont {He}, \citenamefont {Liu}, \citenamefont
  {Shi}, \citenamefont {Lu},\ and\ \citenamefont {Wang}}]{Zhang:2024yfj}%
  \BibitemOpen
  \bibfield  {author} {\bibinfo {author} {\bibfnamefont {Z.}~\bibnamefont
  {Zhang}}, \bibinfo {author} {\bibfnamefont {X.-Y.}\ \bibnamefont {Hu}},
  \bibinfo {author} {\bibfnamefont {G.}~\bibnamefont {He}}, \bibinfo {author}
  {\bibfnamefont {J.}~\bibnamefont {Liu}}, \bibinfo {author} {\bibfnamefont
  {J.-A.}\ \bibnamefont {Shi}}, \bibinfo {author} {\bibfnamefont {B.-N.}\
  \bibnamefont {Lu}},\ and\ \bibinfo {author} {\bibfnamefont {Q.}~\bibnamefont
  {Wang}},\ }\bibfield  {title} {\bibinfo {title} {{Binding of the three-hadron
  $DD*K$ system from the lattice effective field theory}},\ }\href
  {https://doi.org/10.1103/PhysRevD.111.036002} {\bibfield  {journal} {\bibinfo
   {journal} {Phys. Rev. D}\ }\textbf {\bibinfo {volume} {111}},\ \bibinfo
  {pages} {036002} (\bibinfo {year} {2025})},\ \Eprint
  {https://arxiv.org/abs/2409.01325} {arXiv:2409.01325 [hep-ph]} \BibitemShut
  {NoStop}%
\bibitem [{\citenamefont {Pan}\ \emph {et~al.}(2025)\citenamefont {Pan},
  \citenamefont {Lu}, \citenamefont {Hiyama}, \citenamefont {Geng},\ and\
  \citenamefont {Hosaka}}]{Pan:2025xvq}%
  \BibitemOpen
  \bibfield  {author} {\bibinfo {author} {\bibfnamefont {Y.-W.}\ \bibnamefont
  {Pan}}, \bibinfo {author} {\bibfnamefont {J.-X.}\ \bibnamefont {Lu}},
  \bibinfo {author} {\bibfnamefont {E.}~\bibnamefont {Hiyama}}, \bibinfo
  {author} {\bibfnamefont {L.-S.}\ \bibnamefont {Geng}},\ and\ \bibinfo
  {author} {\bibfnamefont {A.}~\bibnamefont {Hosaka}},\ }\bibfield  {title}
  {\bibinfo {title} {{Effect of a repulsive three-body interaction on the
  $DD(*)K$ molecule}},\ }\href {https://doi.org/10.1103/jk8x-qv2p} {\bibfield
  {journal} {\bibinfo  {journal} {Phys. Rev. D}\ }\textbf {\bibinfo {volume}
  {111}},\ \bibinfo {pages} {114006} (\bibinfo {year} {2025})},\ \Eprint
  {https://arxiv.org/abs/2502.00438} {arXiv:2502.00438 [nucl-th]} \BibitemShut
  {NoStop}%
\bibitem [{\citenamefont {Foldy}(1945)}]{Foldy:1945zz}%
  \BibitemOpen
  \bibfield  {author} {\bibinfo {author} {\bibfnamefont {L.~L.}\ \bibnamefont
  {Foldy}},\ }\bibfield  {title} {\bibinfo {title} {{The Multiple Scattering of
  Waves. 1. General Theory of Isotropic Scattering by Randomly Distributed
  Scatterers}},\ }\href {https://doi.org/10.1103/PhysRev.67.107} {\bibfield
  {journal} {\bibinfo  {journal} {Phys. Rev.}\ }\textbf {\bibinfo {volume}
  {67}},\ \bibinfo {pages} {107} (\bibinfo {year} {1945})}\BibitemShut
  {NoStop}%
\bibitem [{\citenamefont {Deloff}(2000)}]{Deloff:1999gc}%
  \BibitemOpen
  \bibfield  {author} {\bibinfo {author} {\bibfnamefont {A.}~\bibnamefont
  {Deloff}},\ }\bibfield  {title} {\bibinfo {title} {{Eta d and K- d zero
  energy scattering: A Faddeev approach}},\ }\href
  {https://doi.org/10.1103/PhysRevC.61.024004} {\bibfield  {journal} {\bibinfo
  {journal} {Phys. Rev. C}\ }\textbf {\bibinfo {volume} {61}},\ \bibinfo
  {pages} {024004} (\bibinfo {year} {2000})}\BibitemShut {NoStop}%
\bibitem [{\citenamefont {Roca}\ and\ \citenamefont
  {Oset}(2010)}]{Roca:2010tf}%
  \BibitemOpen
  \bibfield  {author} {\bibinfo {author} {\bibfnamefont {L.}~\bibnamefont
  {Roca}}\ and\ \bibinfo {author} {\bibfnamefont {E.}~\bibnamefont {Oset}},\
  }\bibfield  {title} {\bibinfo {title} {{A description of the f2(1270),
  rho3(1690), f4(2050), rho5(2350) and f6(2510) resonances as multi-rho(770)
  states}},\ }\href {https://doi.org/10.1103/PhysRevD.82.054013} {\bibfield
  {journal} {\bibinfo  {journal} {Phys. Rev. D}\ }\textbf {\bibinfo {volume}
  {82}},\ \bibinfo {pages} {054013} (\bibinfo {year} {2010})},\ \Eprint
  {https://arxiv.org/abs/1005.0283} {arXiv:1005.0283 [hep-ph]} \BibitemShut
  {NoStop}%
\bibitem [{\citenamefont {Debastiani}\ \emph {et~al.}(2017)\citenamefont
  {Debastiani}, \citenamefont {Dias},\ and\ \citenamefont
  {Oset}}]{Debastiani:2017vhv}%
  \BibitemOpen
  \bibfield  {author} {\bibinfo {author} {\bibfnamefont {V.~R.}\ \bibnamefont
  {Debastiani}}, \bibinfo {author} {\bibfnamefont {J.~M.}\ \bibnamefont
  {Dias}},\ and\ \bibinfo {author} {\bibfnamefont {E.}~\bibnamefont {Oset}},\
  }\bibfield  {title} {\bibinfo {title} {{Study of the $DKK$ and $DK\bar{K}$
  systems}},\ }\href {https://doi.org/10.1103/PhysRevD.96.016014} {\bibfield
  {journal} {\bibinfo  {journal} {Phys. Rev. D}\ }\textbf {\bibinfo {volume}
  {96}},\ \bibinfo {pages} {016014} (\bibinfo {year} {2017})},\ \Eprint
  {https://arxiv.org/abs/1705.09257} {arXiv:1705.09257 [hep-ph]} \BibitemShut
  {NoStop}%
\bibitem [{\citenamefont {Dias}\ \emph {et~al.}(2017)\citenamefont {Dias},
  \citenamefont {Debastiani}, \citenamefont {Roca}, \citenamefont {Sakai},\
  and\ \citenamefont {Oset}}]{Dias:2017miz}%
  \BibitemOpen
  \bibfield  {author} {\bibinfo {author} {\bibfnamefont {J.~M.}\ \bibnamefont
  {Dias}}, \bibinfo {author} {\bibfnamefont {V.~R.}\ \bibnamefont
  {Debastiani}}, \bibinfo {author} {\bibfnamefont {L.}~\bibnamefont {Roca}},
  \bibinfo {author} {\bibfnamefont {S.}~\bibnamefont {Sakai}},\ and\ \bibinfo
  {author} {\bibfnamefont {E.}~\bibnamefont {Oset}},\ }\bibfield  {title}
  {\bibinfo {title} {{On the binding of the $BD\bar{D}$ and $BDD$ systems}},\
  }\href {https://doi.org/10.1103/PhysRevD.96.094007} {\bibfield  {journal}
  {\bibinfo  {journal} {Phys. Rev. D}\ }\textbf {\bibinfo {volume} {96}},\
  \bibinfo {pages} {094007} (\bibinfo {year} {2017})},\ \Eprint
  {https://arxiv.org/abs/1709.01372} {arXiv:1709.01372 [hep-ph]} \BibitemShut
  {NoStop}%
\bibitem [{\citenamefont {Dias}\ \emph {et~al.}(2018)\citenamefont {Dias},
  \citenamefont {Roca},\ and\ \citenamefont {Sakai}}]{Dias:2018iuy}%
  \BibitemOpen
  \bibfield  {author} {\bibinfo {author} {\bibfnamefont {J.~M.}\ \bibnamefont
  {Dias}}, \bibinfo {author} {\bibfnamefont {L.}~\bibnamefont {Roca}},\ and\
  \bibinfo {author} {\bibfnamefont {S.}~\bibnamefont {Sakai}},\ }\bibfield
  {title} {\bibinfo {title} {{Prediction of new states from
  $D^{(*)}B^{(*)}\bar{B}^{(*)}$ three-body interactions}},\ }\href
  {https://doi.org/10.1103/PhysRevD.97.056019} {\bibfield  {journal} {\bibinfo
  {journal} {Phys. Rev. D}\ }\textbf {\bibinfo {volume} {97}},\ \bibinfo
  {pages} {056019} (\bibinfo {year} {2018})},\ \Eprint
  {https://arxiv.org/abs/1801.03504} {arXiv:1801.03504 [hep-ph]} \BibitemShut
  {NoStop}%
\bibitem [{\citenamefont {Martinez~Torres}\ \emph {et~al.}(2020)\citenamefont
  {Martinez~Torres}, \citenamefont {Khemchandani}, \citenamefont {Roca},\ and\
  \citenamefont {Oset}}]{MartinezTorres:2020hus}%
  \BibitemOpen
  \bibfield  {author} {\bibinfo {author} {\bibfnamefont {A.}~\bibnamefont
  {Martinez~Torres}}, \bibinfo {author} {\bibfnamefont {K.~P.}\ \bibnamefont
  {Khemchandani}}, \bibinfo {author} {\bibfnamefont {L.}~\bibnamefont {Roca}},\
  and\ \bibinfo {author} {\bibfnamefont {E.}~\bibnamefont {Oset}},\ }\bibfield
  {title} {\bibinfo {title} {{Few-body systems consisting of mesons}},\ }\href
  {https://doi.org/10.1007/s00601-020-01568-y} {\bibfield  {journal} {\bibinfo
  {journal} {Few Body Syst.}\ }\textbf {\bibinfo {volume} {61}},\ \bibinfo
  {pages} {35} (\bibinfo {year} {2020})},\ \Eprint
  {https://arxiv.org/abs/2005.14357} {arXiv:2005.14357 [nucl-th]} \BibitemShut
  {NoStop}%
\bibitem [{\citenamefont {Guo}\ \emph {et~al.}(2006)\citenamefont {Guo},
  \citenamefont {Shen}, \citenamefont {Chiang}, \citenamefont {Ping},\ and\
  \citenamefont {Zou}}]{Guo:2006fu}%
  \BibitemOpen
  \bibfield  {author} {\bibinfo {author} {\bibfnamefont {F.-K.}\ \bibnamefont
  {Guo}}, \bibinfo {author} {\bibfnamefont {P.-N.}\ \bibnamefont {Shen}},
  \bibinfo {author} {\bibfnamefont {H.-C.}\ \bibnamefont {Chiang}}, \bibinfo
  {author} {\bibfnamefont {R.-G.}\ \bibnamefont {Ping}},\ and\ \bibinfo
  {author} {\bibfnamefont {B.-S.}\ \bibnamefont {Zou}},\ }\bibfield  {title}
  {\bibinfo {title} {{Dynamically generated 0+ heavy mesons in a heavy chiral
  unitary approach}},\ }\href {https://doi.org/10.1016/j.physletb.2006.08.064}
  {\bibfield  {journal} {\bibinfo  {journal} {Phys. Lett. B}\ }\textbf
  {\bibinfo {volume} {641}},\ \bibinfo {pages} {278} (\bibinfo {year}
  {2006})},\ \Eprint {https://arxiv.org/abs/hep-ph/0603072}
  {arXiv:hep-ph/0603072} \BibitemShut {NoStop}%
\bibitem [{\citenamefont {Guo}\ \emph {et~al.}(2007)\citenamefont {Guo},
  \citenamefont {Shen},\ and\ \citenamefont {Chiang}}]{Guo:2006rp}%
  \BibitemOpen
  \bibfield  {author} {\bibinfo {author} {\bibfnamefont {F.-K.}\ \bibnamefont
  {Guo}}, \bibinfo {author} {\bibfnamefont {P.-N.}\ \bibnamefont {Shen}},\ and\
  \bibinfo {author} {\bibfnamefont {H.-C.}\ \bibnamefont {Chiang}},\ }\bibfield
   {title} {\bibinfo {title} {{Dynamically generated 1+ heavy mesons}},\ }\href
  {https://doi.org/10.1016/j.physletb.2007.01.050} {\bibfield  {journal}
  {\bibinfo  {journal} {Phys. Lett. B}\ }\textbf {\bibinfo {volume} {647}},\
  \bibinfo {pages} {133} (\bibinfo {year} {2007})},\ \Eprint
  {https://arxiv.org/abs/hep-ph/0610008} {arXiv:hep-ph/0610008} \BibitemShut
  {NoStop}%
\bibitem [{\citenamefont {Ikeno}\ and\ \citenamefont
  {Oset}(2025)}]{Ikeno:2025bsx}%
  \BibitemOpen
  \bibfield  {author} {\bibinfo {author} {\bibfnamefont {N.}~\bibnamefont
  {Ikeno}}\ and\ \bibinfo {author} {\bibfnamefont {E.}~\bibnamefont {Oset}},\
  }\bibfield  {title} {\bibinfo {title} {{Correlation function for the
  $nD_{s0}*(2317)$ interaction and the issue of elastic unitarity}},\ }\href
  {https://doi.org/10.1103/bb31-rdjb} {\bibfield  {journal} {\bibinfo
  {journal} {Phys. Rev. D}\ }\textbf {\bibinfo {volume} {112}},\ \bibinfo
  {pages} {094019} (\bibinfo {year} {2025})},\ \Eprint
  {https://arxiv.org/abs/2507.16367} {arXiv:2507.16367 [hep-ph]} \BibitemShut
  {NoStop}%
\bibitem [{\citenamefont {Yamagata-Sekihara}\ \emph {et~al.}(2010)\citenamefont
  {Yamagata-Sekihara}, \citenamefont {Roca},\ and\ \citenamefont
  {Oset}}]{Yamagata-Sekihara:2010muv}%
  \BibitemOpen
  \bibfield  {author} {\bibinfo {author} {\bibfnamefont {J.}~\bibnamefont
  {Yamagata-Sekihara}}, \bibinfo {author} {\bibfnamefont {L.}~\bibnamefont
  {Roca}},\ and\ \bibinfo {author} {\bibfnamefont {E.}~\bibnamefont {Oset}},\
  }\bibfield  {title} {\bibinfo {title} {{On the nature of the $K^*_2(1430)$,
  $K^*_3(1780)$, $K^*_4(2045)$, $K^*_5(2380)$ and $K^*6$ as $K^*$ -
  multi-$\rho$ states}},\ }\href {https://doi.org/10.1103/PhysRevD.82.094017}
  {\bibfield  {journal} {\bibinfo  {journal} {Phys. Rev. D}\ }\textbf {\bibinfo
  {volume} {82}},\ \bibinfo {pages} {094017} (\bibinfo {year} {2010})},\
  \bibinfo {note} {[Erratum: Phys.Rev.D 85, 119905 (2012)]},\ \Eprint
  {https://arxiv.org/abs/1010.0525} {arXiv:1010.0525 [hep-ph]} \BibitemShut
  {NoStop}%
\bibitem [{\citenamefont {Brand{\~a}o}\ \emph {et~al.}(2026)\citenamefont
  {Brand{\~a}o}, \citenamefont {Agat{\~a}o}, \citenamefont {Abreu},
  \citenamefont {Khemchandani},\ and\ \citenamefont
  {Mart{\'\i}nez~Torres}}]{Brandao:2025cli}%
  \BibitemOpen
  \bibfield  {author} {\bibinfo {author} {\bibfnamefont {P.}~\bibnamefont
  {Brand{\~a}o}}, \bibinfo {author} {\bibfnamefont {B.}~\bibnamefont
  {Agat{\~a}o}}, \bibinfo {author} {\bibfnamefont {L.~M.}\ \bibnamefont
  {Abreu}}, \bibinfo {author} {\bibfnamefont {K.~P.}\ \bibnamefont
  {Khemchandani}},\ and\ \bibinfo {author} {\bibfnamefont {A.}~\bibnamefont
  {Mart{\'\i}nez~Torres}},\ }\bibfield  {title} {\bibinfo {title} {{$D^*\pi$
  interaction from the lineshape of $D_1(2420)$ in $B$-decays}},\ }\href
  {https://doi.org/10.1016/j.physletb.2026.140527} {\bibfield  {journal}
  {\bibinfo  {journal} {Phys. Lett. B}\ }\textbf {\bibinfo {volume} {878}},\
  \bibinfo {pages} {140527} (\bibinfo {year} {2026})},\ \Eprint
  {https://arxiv.org/abs/2512.24370} {arXiv:2512.24370 [hep-ph]} \BibitemShut
  {NoStop}%
\bibitem [{\citenamefont {Yamagata-Sekihara}\ \emph {et~al.}(2011)\citenamefont
  {Yamagata-Sekihara}, \citenamefont {Nieves},\ and\ \citenamefont
  {Oset}}]{Yamagata-Sekihara:2010kpd}%
  \BibitemOpen
  \bibfield  {author} {\bibinfo {author} {\bibfnamefont {J.}~\bibnamefont
  {Yamagata-Sekihara}}, \bibinfo {author} {\bibfnamefont {J.}~\bibnamefont
  {Nieves}},\ and\ \bibinfo {author} {\bibfnamefont {E.}~\bibnamefont {Oset}},\
  }\bibfield  {title} {\bibinfo {title} {{Couplings in coupled channels versus
  wave functions in the case of resonances: application to the two
  $\Lambda(1405)$ states}},\ }\href
  {https://doi.org/10.1103/PhysRevD.83.014003} {\bibfield  {journal} {\bibinfo
  {journal} {Phys. Rev. D}\ }\textbf {\bibinfo {volume} {83}},\ \bibinfo
  {pages} {014003} (\bibinfo {year} {2011})},\ \Eprint
  {https://arxiv.org/abs/1007.3923} {arXiv:1007.3923 [hep-ph]} \BibitemShut
  {NoStop}%
\bibitem [{\citenamefont {Encarnaci{\'o}n}\ \emph {et~al.}(2026)\citenamefont
  {Encarnaci{\'o}n}, \citenamefont {Feijoo},\ and\ \citenamefont
  {Oset}}]{Encarnacion:2026zas}%
  \BibitemOpen
  \bibfield  {author} {\bibinfo {author} {\bibfnamefont {P.}~\bibnamefont
  {Encarnaci{\'o}n}}, \bibinfo {author} {\bibfnamefont {A.}~\bibnamefont
  {Feijoo}},\ and\ \bibinfo {author} {\bibfnamefont {E.}~\bibnamefont {Oset}},\
  }\bibfield  {title} {\bibinfo {title} {{Scattering observables and
  correlation function for $p f1(1285)$ revisited}},\ }\href
  {https://doi.org/10.1103/kk4c-xv4d} {\bibfield  {journal} {\bibinfo
  {journal} {Phys. Rev. D}\ }\textbf {\bibinfo {volume} {113}},\ \bibinfo
  {pages} {L111502} (\bibinfo {year} {2026})},\ \Eprint
  {https://arxiv.org/abs/2603.09852} {arXiv:2603.09852 [hep-ph]} \BibitemShut
  {NoStop}%
\bibitem [{\citenamefont {Agat{\~a}o}\ \emph {et~al.}(2025)\citenamefont
  {Agat{\~a}o}, \citenamefont {Brand{\~a}o}, \citenamefont
  {Mart{\'\i}nez~Torres}, \citenamefont {Khemchandani}, \citenamefont {Abreu},\
  and\ \citenamefont {Oset}}]{Agatao:2025ckp}%
  \BibitemOpen
  \bibfield  {author} {\bibinfo {author} {\bibfnamefont {B.}~\bibnamefont
  {Agat{\~a}o}}, \bibinfo {author} {\bibfnamefont {P.}~\bibnamefont
  {Brand{\~a}o}}, \bibinfo {author} {\bibfnamefont {A.}~\bibnamefont
  {Mart{\'\i}nez~Torres}}, \bibinfo {author} {\bibfnamefont {K.~P.}\
  \bibnamefont {Khemchandani}}, \bibinfo {author} {\bibfnamefont {L.~M.}\
  \bibnamefont {Abreu}},\ and\ \bibinfo {author} {\bibfnamefont
  {E.}~\bibnamefont {Oset}},\ }\bibfield  {title} {\bibinfo {title}
  {{Correlation functions for $n\,\bar{D}_{s1}(2460)$ and
  $n\,\bar{D}_{s1}(2536)$}},\ }\href
  {https://doi.org/10.1140/epjc/s10052-025-14838-y} {\bibfield  {journal}
  {\bibinfo  {journal} {Eur. Phys. J. C}\ }\textbf {\bibinfo {volume} {85}},\
  \bibinfo {pages} {1136} (\bibinfo {year} {2025})},\ \Eprint
  {https://arxiv.org/abs/2508.05825} {arXiv:2508.05825 [hep-ph]} \BibitemShut
  {NoStop}%
\end{thebibliography}%

\end{document}